# Observation of liquid-solid transition of nanoconfined water at ambient temperature

Wentian Zheng[1,5*], Shichen Zhang[1*], Jian Jiang[2,4*], Yipeng He[1], Rainer Stöhr[6,7], Andrej Denisenko[6,7], Jörg Wrachtrup[6,7], Xiao Cheng Zeng[2,3†], Ke Bian[1,9†], En-Ge Wang[1,8,9,10†], Ying Jiang[1,8,9,11†]

[1]*International Center for Quantum Materials, School of Physics, 100871, Peking University, Beijing, 100871, P. R. China*

[2]*Department of Material Science and Engineering, City University of Hong Kong, Hong Kong 999077, P. R. China*

[3]*Hong Kong Institute for Clean Energy, City University of Hong Kong, Hong Kong 9990777, P. R. China*

[4]*Shenzhen Research Institute, City University of Hong Kong, Shenzhen 518057, P. R. China*

[5]*Department of Physics and Astronomy, Rice University, Houston 77005, TX, USA*

[6]*3rd Institute of Physics, University of Stuttgart and Institute for Quantum Science and Technology (IQST), Stuttgart, 70569, Germany*

[7]*Max Planck Institute for Solid State Research, Stuttgart, 70569, Germany*

[8]*Collaborative Innovation Center of Quantum Matter, Beijing 100871, P. R. China*

†E-mail: bian.k@pku.edu.cn (K.B.); xzeng26@cityu.edu.hk (X.-C.Z.); egwang@pku.edu.cn (E.-G.W.); yjiang@pku.edu.cn (Y.J.)
*These authors contribute equally to this work.

[9]*Interdisciplinary Institute of Light-Element Quantum Materials and Research Center for Light-Element Advanced Materials, Peking University, Beijing, 100871, P. R. China*

[10]*Tsientang Institute for Advanced Study, Zhejiang, 310024, P. R. China*

[11]*New Cornerstone Science Laboratory, Peking University, Beijing 100871, P. R. China*

**Nanoconfined water plays an indispensable role in various phenomena in biology, chemistry, and engineering. It exhibits many abnormal properties[1-3] compared to bulk water, especially under strong confinement. However, the origin of those anomalies is still elusive due to the lack of structural information on hydrogen-bonding networks. Considering the inhomogeneity of the nanocavity and the tiny amount of water molecules, conventional optical spectroscopies and nuclear magnetic resonance (NMR) fail to realize the structure analysis of nanoconfined water. Here, we addressed this issue by combining scanning probe microscopy (SPM) with advanced quantum sensing[4,5] (QS) based on an atomic-size quantum sensor like nitrogen-vacancy (NV) center in diamond, which can apply the nanoscale-NMR[6,7] for characterizing both the dynamics[8] and structure[9,10] of confined water at ambient conditions. We built a two-dimensional (2D) nanoconfined water system with a hexagonal-boron nitride (hBN) flake and a hydrophilic diamond surface. By using the SPM tip to measure the confinement size precisely, we observed a critical confinement size of ~2 nm, below which the**

**water diffusion was significantly suppressed and the hydrogen-bonding network of water showed an ordered structure. Meanwhile, molecular dynamics (MD) simulation revealed a solid-like water contact layer on the diamond surface under strong confinement, which also reproduced the measured nanoscale-NMR spectra and confirmed the liquid-solid phase transition observed in the experiments. Notably, with this new SPM-QS platform, our results showed a promising way to elucidate the abnormal properties of nanoconfined water in future applications.**

Confinement is a universal phenomenon that often leads to the emergence of new phases of matter by limiting motional degrees of freedom and breaking spatial symmetries, as described by Landau's theory[11]. Nanoconfined water stands out among various confined systems, drawing considerable interest due to its critical role in numerous natural and technological processes, including biological water[12-14], water purification and desalination[15], water collection[16,17], and energy harvesting[18]. To date, a variety of abnormal properties in nanoconfined water have been reported. For example, water confined within low-dimensional nanochannels engineered by graphene or carbon nanotubes has shown anomalously low dielectric constant[2], ultra-fast molecular transportation[1,19,20], and ferroelectricity[3,21]. To clarify the origin of those anomalies, it is crucial to reveal the structure of the hydrogen-bonding water network under nano- or even sub-nano-confinement, which has been extensively investigated using density function theory (DFT) calculations and molecular dynamics (MD) simulations[22-26]. It was suggested that the structural order of water under strong confinement could be greatly enhanced.

Despite the massive theoretical efforts, experimentally characterizing the structure of nanoconfined water has remained a grand challenge, due to the nanoscale inhomogeneity of the nanocavity and the tiny amount of water molecules. For example, scanning probe microscopy (SPM) was capable of visualizing the ice structures with atomic resolution[27,28], but failed to directly address the water molecules within the nanocavity because of the limited sensitivity for interfacial species. Meanwhile, conventional spectroscopies such as nuclear magnetic resonance (NMR)[29,30], optical spectroscopies[31-33], and X-ray diffraction (XRD)[34] were powerful tools for revealing the atomic-level structures of materials. Still, they could not obtain local structure information of nanoconfined species due to the averaging effect. Recently, the atomic-size nitrogen-vacancy (NV) center in diamond was demonstrated as a remarkable quantum sensor with nanometer resolution[6,7,35]. Using shallow NV centers in proximity to the diamond surface, the NMR measurements can be realized with sensitivity even up to single nuclear spins[36,37], allowing the identification of multiple nuclear species[7] and the analysis of intra-molecule structures by chemical shifts[9] at the nanoscale.

Here we utilized a new type of qPlus-SPM[38] compatible with the NV-based quantum sensing (QS)[39] to investigate the nanoconfined water under ambient conditions, which not only enabled nanoscale NMR measurements for analyzing the structure and dynamics of the nanoconfined water but also provided topographic imaging for evaluating the confinement size. In this work, water molecules were confined by a two-dimensional (2D) nanocavity consisting of a hexagonal-boron nitride (hBN) flake and a hydrophilic diamond surface, where the confinement size could be finely tuned. Leveraging the NV-based nanoscale NMR (NV-NMR), we observed the diffusion rate of water decreased rapidly along with the increased confinement strength at a critical confinement size of ~2 nm. Meanwhile, we found the appearance of fine structures in the NMR spectra as direct evidence of ordered hydrogen-bonding networks. The determined critical confinement size was consistent with that in previous works for observing abnormal properties of confined water[1,2,19,40,41]. To obtain further insight into this phase transition, we performed molecular dynamics (MD) simulation and revealed considerable suppression of translational and rotational motions of water molecules near the critical confinement size of ~2 nm, followed by the formation of a solid-like contact layer of water on the diamond surface under strong confinement. Besides, the hydrogen-bonding structures of such a solid-like layer reproduced the experimental NMR spectra. These results demonstrated the combination of SPM and quantum sensors such as NV centers offers a promising platform to investigate and elucidate the anomalous properties of confined water in future applications.

Our experimental setup is illustrated in Fig. 1a, with the NV-NMR technology integrated into a home-built SPM at ambient conditions (see Methods for details). Single shallow NV centers were located at a depth of <10 nm below a hydrophilic diamond surface after acid treatments. By introducing the boron-doped sacrifice layers before ion implantations[42] (see Methods for details), the spin coherence time of NV centers was enhanced to detect nuclear spins outside the diamond surface. The quantum states of single-NV were addressed by a 532-nm laser under a bias magnetic field ($B_z$ ~300 Gauss in this work) and coherently manipulated by a train of microwave pulses for NV-NMR. To construct the nanoconfinement, an hBN flake was exfoliated and transferred onto the diamond surface using a dry transfer technique[43], with controlled humidity and temperature to tune the thickness of water layers (see Methods for details). In this configuration, water molecules adsorbed on the diamond surface were squeezed by the van der Waals interactions between hBN and the diamond surface. Simultaneously, we used qPlus-based SPM[38] to characterize the surface topography and confinement size with a precision of <1 nm.

Fig. 1d and 1e show the same region of the diamond surface imaged by qPlus-SPM before and after being covered by an hBN flake, respectively. The position of a single NV center is marked by an orange arrow, determined by the quenching effect of a conductive SPM tip (Supplementary Text 1 and Supplementary Fig. S1). Owing to a series of procedures including polishing, implantation, and etching to produce shallow NV centers[42], residual nano-diamonds with sizes below 5 nm remained on the surface (denoted by the white dashed circles in Fig. 1d and 1e). Those nano-diamonds were stable and survived after tri-acid boiling without changing their heights and distributions, thus providing reference landmarks for measuring the thickness of water layers (as shown in Fig. 1b and 1c). The schematics in Fig. 1b and 1c illustrate the principle for measuring the thickness of the confined water layer. Specifically, they depict the height difference between the nano-diamonds and the water layer without ($d_1$) and with ($d_2$) hBN confinement. Then the confinement size, $d_{\text{confine}}$, is calculated as $d_{\text{confine}}=d_{\text{water}}-(d_2-d_1)$, where $(d_2-d_1)$ represents the reduced thickness of the water layer during the transfer of hBN flakes. And $d_{\text{water}}$ is the initial thickness of the adsorbed water layer of the open system, which can be measured by comparing the height difference in topographic imaging for the open system to the confinement structure without water molecules inside (Supplementary Text 2). Fig. 1f and 1g show the extracted $d_1$ and $d_2$ according to the height profiles along the black dashed lines in Fig. 1d and 1e (for more examples, see Supplementary Text 2 and Supplementary Fig. S2).

At the same time, proton spins of water molecules were sensed by NV-NMR methods[6,8,10] (Supplementary Text. 3 and Supplementary Fig. 3). In this case, correlation spectroscopy[8] was applied, resembling the nuclear free induction decay (FID) in conventional NMR. This measurement doesn't require the polarization of proton spins and the decay time is primarily limited by their diffusion dynamics on diamond surface at room temperature. Fig. 2a shows an oscillating signal with an overall decayed envelope for the water molecules without confinement. The characteristic time constant $T_{\text{corr}}$=22.1±4.77 μs was fitted by $S(\tilde{\tau})=S_0 \cdot \cos(2\pi f_L \tilde{\tau}+\varphi_0) \cdot e^{-\tilde{\tau}/T_{\text{corr}}}$, where $f_L$ denoted the oscillating frequency. After performing a fast Fourier transformation (FFT), a resonance peak appeared at protons' Larmor frequency (1.278 MHz under $B_z$~302 Gauss) with a linewidth of 14.2 kHz fitted by a Lorentz function. In liquid-state NMR, $T_{\text{corr}}$ directly reflects the diffusivity of water molecules, which continuously move into/out of the detection volume of ~5 nm³ of a single shallow NV[8]. Accordingly, we extracted the diffusion coefficient of water molecules in Fig. 2a (Supplementary Text 4) on the diamond surface as $D_{water}$=(0.388±0.131)×10$^{-12}$ m$^2$s$^{-1}$, about three orders of magnitude smaller than that in bulk water ($D_{water}$=2.3×10$^{-9}$ m$^2$s$^{-1}$ at 300 K), possibly due to the strong interaction between hydrophilic diamond surface and water molecules. In Fig. 2b, $T_{corr}$ increased to 66.9±13.31 µs with a narrower linewidth of 4.7 kHz after transferring hBN onto the diamond surface at a confinement size of 2.3±0.4 nm, leading to a smaller diffusion coefficient $D_{water}$=(0.127±0.041)×10$^{-12}$ m$^2$s$^{-1}$. We attributed it to the hindered motions of water molecules under the confinement. It's noted that an under-sampling protocol (see Methods and Supplementary Fig. S4 for details) was adopted to guarantee enough signal-to-noise ratio, which transformed the protons' Larmor frequency to $f_L^* = f_L - nf_s$ (where $f_s$ is the sampling frequency and $n$ is the quotient of $f_L/f_s$) without changing the spectral information.

When the confinement size was further decreased to 1.0±0.3 nm, an additional beating feature emerged in the correlation spectrum (Fig. 2c) and we obtained a $T_{corr}$ up to 183.6±33.03 µs, fitted by a sum of cosine functions with multiple frequencies and an exponentially decayed envelope $S(\tilde{\tau}) = e^{-\tilde{\tau}/T_{corr}} \cdot \sum_i S_i \cdot \cos(2\pi f_i \tilde{\tau} + \varphi_i)$. Correspondingly, the diffusion coefficient was calculated as $D_{water}$=(0.046±0.013)×10$^{-12}$ m$^2$s$^{-1}$. It's worth noting that we only monitored the protons' lifetime up to ~200 µs with a total integration time of about 10 hours, which was limited by the absolute signal-to-noise ratio of single NV center and might lead to an overestimation of $D_{water}$ in Fig. 2c. Notably, a fine structure with multiple peaks appeared in FFT (with the fitted frequencies as $f_1$=1.246 MHz, $f_2$=1.266 MHz, $f_3$=1.276 MHz, $f_4$=1.291 MHz). The observed frequency splitting of 10~25 kHz near the protons' Larmor frequency suggested the magnetic dipolar interactions between intra- or intermolecular proton spins[10] (Supplementary Text 5). Considering a pair of protons, the magnetic dipolar interaction is formulated as: $H_p \propto (1-3\cos^2\theta)\cdot(3\vec{I}_1\vec{I}_2-\boldsymbol{I}_1\boldsymbol{I}_2)/d^3$, where $d$ is the distance between these two protons, and $\theta$ is the angle of the proton pairs relative to $B_z$. Since on the acid-treated diamond surface, the coverage of the hydroxyl group was estimated to be lower than 20%[44,45], the characteristic distance between the hydroxyl was random and should be larger than the lattice constant of the diamond. Therefore, we excluded the contribution of carbon-hydroxyl in Fig. 2c and attributed the splitting to the intra- and intermolecular interactions of the hydrogen-bonding networks of water molecules. In the liquid phase, since $\theta$ and $d$ are always randomly distributed both in space and time due to the fast dynamics of water molecules at room temperature, the splitting effect from dipolar interaction between protons will be averaged[8]. In contrast, for a solid-like phase with ordered structures (such as ice), the diffusion/rotation of the water molecule is almost frozen and $\theta, d$ can only be a set of discrete values, thus leading to the discrete peak features near the protons' spin resonance[10].

By applying systematic measurements to nanoconfined water with $d_{\text{confine}}$ ranging from 1 to 6 nm, an overall trend of liquid-solid phase transition was observed as shown in the diagram of Fig. 3a. The diffusion coefficient $D_{\text{water}}=(0.388\pm0.131)\times10^{-12}$ m²s⁻¹ without confinement (denoted as "open system" in the diagram) was listed as reference (black data points in Fig. 3a). As long as the water molecules were confined, $D_{\text{water}}$ continuously decreased (red data points) along with decreasing $d_{\text{confine}}$. Notably, $D_{\text{water}}$ suddenly dropped at $d_{\text{confine}}\sim 2$ nm and then leveled off at the value of $0.046\times10^{-12}$ m²s⁻ [1], which was nearly an order of magnitude smaller than that of the open system. When $d_{\text{confine}}$ is smaller than 2 nm, the splitting features in nanoscale NMR spectra appeared (denoted as blue diamonds in Fig. 3a). The whole trend of $D_{\text{water}}$ resembled a step function with a transition point at ~2 nm, where only five layers of water molecules were accommodated in this nanocavity, suggesting the occurrence of liquid-solid phase transition.

To gain deeper insight into the microscopic origins of this phase transition, we performed MD simulations for nanoconfined water (see Methods and Supplementary Text 6 for details). We considered the oxidized diamond (100) surface terminated with different hydrophilic functional groups (Fig. 4a) including ether-like, ketone-like, and alcohol-like groups as previously studied by X-ray photoelectron spectroscopy (XPS)[44]. Both the translational ($D_t$) and rotational ($D_r$) coefficients were calculated through the statistical trajectories of water molecules with $d_{\text{confine}}$ ranging from 1 nm to 6 nm, as shown in Fig. 3b. $D_t=(5.11\pm1.44)\times10^{-10}$ m²/s was obtained at $d_{\text{confine}}=6$ nm and decreased to $(0.20\pm0.09)\times10^{-10}$ m²/s with reduced $d_{\text{confine}}$. In the log-log plot, the drop rate of $D_t$ changed from a smaller slope (blue dashed line) to a much steeper slope (red dashed line) at the confinement size of 2 nm, which was consistent with the experimental results shown in Fig. 3a. A similar decrease of $D_r$ was shown in Fig. 3b, suggesting a concerted suppression of both translational and rotational degrees of freedom of water molecules due to the increased confinement strength.

Focusing on the contact water layer on a diamond surface, the hydrogen-bonding network is globally disordered at $d_{\text{confine}}=6$ nm, with both oxygen and hydrogen atoms being randomly distributed in real space (Fig. 4b). By contrast, a four-fold symmetric spatial distribution of hydrogen bonds was observed at $d_{\text{confine}}$=1 nm (Fig. 4c), following the symmetry of the diamond (100) surface (Fig. 4a). We attributed this disorder-to-order transition of the contact-layer water molecules to the delicate competition between the interfacial water-surface and bulk water-water interactions. Furthermore, we also validated the results by using *ab initio* molecular dynamics (AIMD) simulations, which demonstrated a consistently enhanced ordering of hydrogen bonds at $d_{\text{confine}}$ of 1 nm (see Supplementary Fig. S5).

Finally, we simulated the NMR spectra using the structures of the contact layer obtained by MD simulations. We extracted clusters consisting of 5 water molecules, i.e., 10 protons, (for example, one cluster was marked by the dashed circle in Fig. 4c) as the representative structures, since these clusters serve as the fundamental structural units with tetrahedral hydrogen bonds and include the dipolar interactions both from the nearest-neighbor and next-nearest neighbor protons. We statistically summarized the representative 5-water cluster of all snapshots in the simulation time window and classified them according to the spatial distribution of both oxygen and hydrogen atoms in the lower panel of Fig. 4c. Then we averaged all the 5-water clusters for each category and used them to fit the NV-NMR spectra (Supplementary Text 7 and Supplementary Fig. S6). Fig. 4d shows the calculated NMR (grey curve) from one representative cluster using the experimental parameters in Fig. 2c, which reproduced the splitting features of the resonance peaks in experimental data (red curve). The overall shape of the curve was determined by the filter function of the multi-π pulses in the XY8-k sequence, while the positions and intensities of the peaks point to the specific hydrogen-bonding configuration. The resonance peaks of the calculated NMR spectrum showed a slight overall shift compared to the measured one, probably due to the inaccuracy of determining $B_z$ from the experimental data. Interestingly, we noted that both the shape and peak distribution were extremely sensitive to the position of oxygen/hydrogen atoms and the orientation of water molecules (more examples in Supplementary Fig. S7), indicating the potential of NV-NMR to analyze hydrogen-bonding networks even down to the atomic scale[46]. For comparison, an NMR spectrum based on the disordered hydrogen bonds (blue dashed curve) in Fig. 4b was also calculated, revealing a broadened resonance peak due to the randomly distributed intra- and intermolecular interacting strength. These results further confirm the enhanced ordering of hydrogen bonds and solid-like phase of water molecules under strong confinement.

In summary, our results provide compelling evidence for the liquid-solid phase transition of nanoconfined water under ambient conditions. The combined SPM with NV-NMR techniques can be generalized to other low-dimensional confined systems, such as nanopores, nanotubes, and 2D materials, as long as the thickness of the confining wall is less than 5 nm. Further improvements in the signal-to-noise ratio and spectral resolution can be realized either by structured photonics on the diamond surface[47] or nuclear-based quantum memories[9]. Meanwhile, by rotating $B_z$, a full set of the structural information of nuclear spins can be obtained[48] by analyzing the magnetic dipolar interactions. We emphasize that the observed critical confinement size (~2 nm) coincides with that of the emergence of the anomalies of nanoconfined water. Our findings of the ordered structure of confined water would resolve the longstanding debates and uncertainties regarding the origins of various anomalous properties. For example, the water transport could become collective in the ice-like phase, leading to superlubricity[49] and drastically enhanced transport rates through low-dimensional channels, such as graphene[20] or carbon nanotubes[1,19]. In addition, the stabilized and ordered electric dipoles observed here also shed light on the microscopic origins of ferroelectricity and the anomalous dielectric constant of confined water[2,3,21].

**Acknowledgments**

The authors thank Lin Yang and Chuanli Yu from Peking University (PKU) for their help with 2D material transfer and fabrications, and we thank Nan Zhao from Beijing Computational Science Research Center (CSRC) for meaningful guidance and discussions on the NMR simulations. This work was supported by the National Key R&D Program under Grant No. 2021YFA1400500, the Program under Grant No. 2023ZD0301300, the National Natural Science Foundation of China under Grant Nos 11888101, 21725302, 12474160, U22A20260, and 12250001, the Strategic Priority Research Program of the Chinese Academy of Sciences under Grant Nos XDB28000000 and the Beijing Municipal Science & Technology Commission under Grant No. Z231100006623009. W.Z. acknowledges the China Postdoctoral Science Foundation under Grant No. 2022M710235. Y.J. acknowledges the New Cornerstone Science Foundation through the New Cornerstone Investigator Program and the XPLORER PRIZE and the Beijing Outstanding Young Scientist Program under Grant No. JWZQ20240101002. X.-C.Z. acknowledges the support from the Hong Kong Global STEM Professorship Scheme and the Research Grants Council of Hong Kong (GRF Grant No. 11204123). J.J. acknowledges the funding support of the National Natural Science Foundation of China (Grant No. 22303072). R.S., A.D., and J.W. acknowledges the BMBF via project QSENS grant 03ZA1110HA and the DFG via FOR 2724 and GRK 2642 as well as WR 28/34-1.

**Author contributions:** Y.J. and E.-G.W. supervised the project. W.Z., S.Z., K.B., and Y.J. designed the experiment. R.S. and A.D. grew diamond chips and fabricated shallow NVs. S.Z. and W.Z. transfer the hBN flakes and fabricate the hBN-diamond structure. W.Z. and S.Z. performed experiments and data acquisition. W.Z., S.Z., J.J., K.B., Y.H., J.W., X.-C.Z., Y.J., and E.-G.W. performed experimental data analysis and interpretation. J.J. and X.-C.Z. performed the MD simulations. W.Z., S.Z., J.J., and K.B. performed the NMR simulations. W.Z., K.B., S.Z., J.J., X.-C.Z., Y.J., and E.-G.W. wrote the manuscript with the inputs from all other authors. All the authors commented on the final manuscript.

**Competing Interests Statement**

The authors declare no competing interests.

**Figure Captions:**

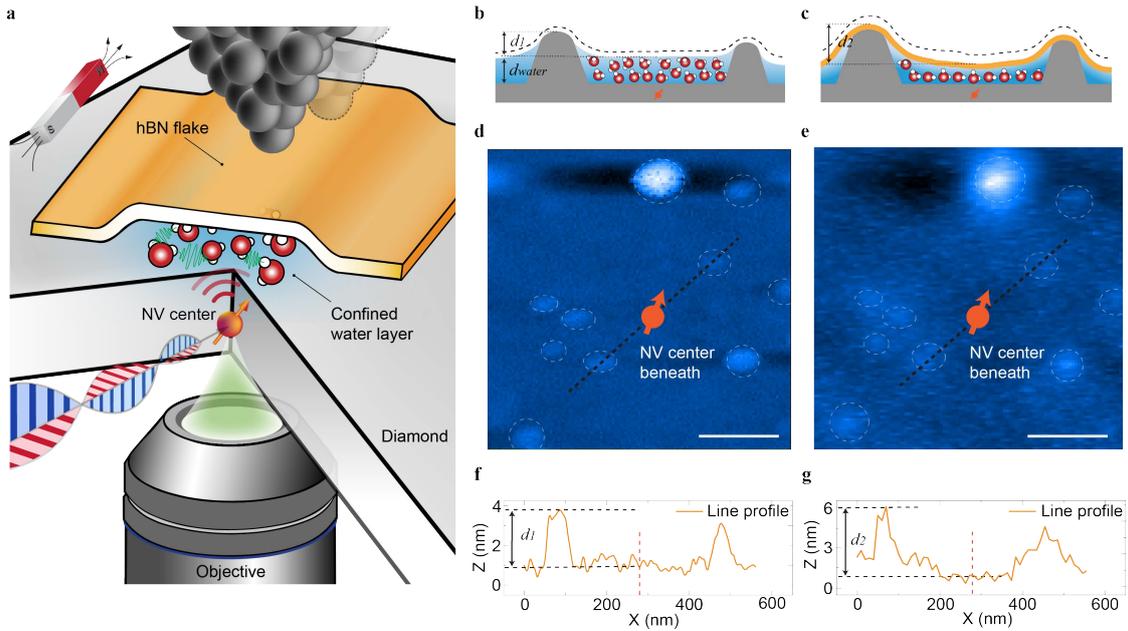

**Fig. 1 | Schematics of the experiment setup and AFM measurements for the confinement size. (a)** Schematics of the experimental setup. All measurements were carried out inside a homemade SPM compatible with the advanced QS technologies at ambient conditions. A single NV center (marked as a red arrow) in the diamond membrane (grey chip) was optically addressed by an objective and manipulated by a microwave sequence. An exfoliated hBN flake (orange sheet) was transferred to the diamond surface to construct a 2D nanoconfinement for water molecules. A tungsten AFM tip was adopted for surface characterization and a permanent magnet was aligned along the axis of NV. **(b)** and **(c)** The schematic cross sections without (b) and with (c) hBN confinement, where the water molecules were squeezed by the van der Waals interactions. The black dashed lines denote the corrugations during the AFM scan. $d_1$ and $d_2$ represent the distances from the top of nano-diamonds to the water layer without (b) and with (c) hBN confinement, while $d_{water}$ represents the original thickness of the water layer on the open system. **(d)** and **(e)** Topographic images measured by AFM without (d) and with (e) hBN confinement. The residual nano-diamonds are marked by

the white dashed circles, where the position of the NV was determined by the quenching effect of a tungsten tip (Supplementary Text 1 and Supplementary Fig. S1). Scale bars, 200 nm. **(f)** and **(g)** Line profiles (orange curve) along the black dashed lines in (d) and (e), showing the measured $d_1$ and $d_2$. The position of NV center is denoted as the vertical red dashed line in (f) and (g).

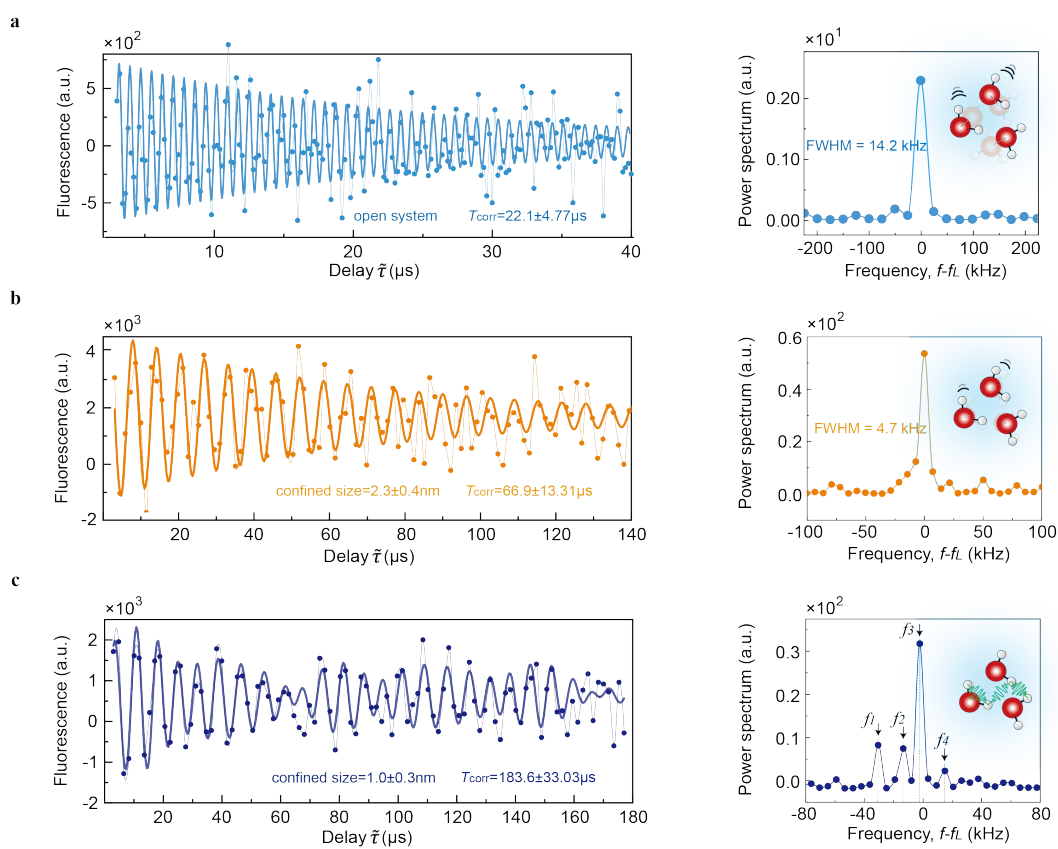

**Fig. 2 | NV-NMR spectroscopy of confined water at different confinement sizes. (a)** Left: The correlation spectrum on the water molecules without hBN confinement, with a delay time up of to 40 μs. This data was fitted by a single-frequency cosine function enveloped by an exponential decay (cyan curve). The obtained time constant was 22.1±4.77 μs. Right: The corresponding FFT analysis. A Lorentz-shaped resonance peak is centered at the proton's Larmor frequency $f_L$=1.278 MHz under an external field of 302 Gauss. The full width at half maximum (FWHM) is 14.2 kHz. **(b)** Left: The

correlation spectrum under a weak confinement was carried out by an under-sampling algorithm which extended the delay time up to 140 μs. This data was fitted with a time constant of 66.9±13.31 μs (orange curve). The confinement size was 2.3±0.4 nm. Right: The corresponding FFT analysis shows a resonance peak at the same frequency in (a) but with a narrowed FWHM of 4.7 kHz. **(c)** Left: The correlation spectrum under a strong confinement was carried out by an under-sampling algorithm which shows a beating feature along with the delay time. This data was fitted by a multi-frequency cosine function with an overall exponential decay (dark-blue curve). The obtained time constant is 183.6±33.03 μs. The confinement size is 1.0±0.3 nm. Right: The corresponding FFT analysis shows the multi frequencies $f_{1,2,3,4}$ near the resonance frequency $f_L$, with a splitting strength varied from 10 kHz to 25 kHz. The insets of the right panels in (a)-(c) show the schematics of hindered motions and stabilized intermolecular interaction of water molecules under confinement.

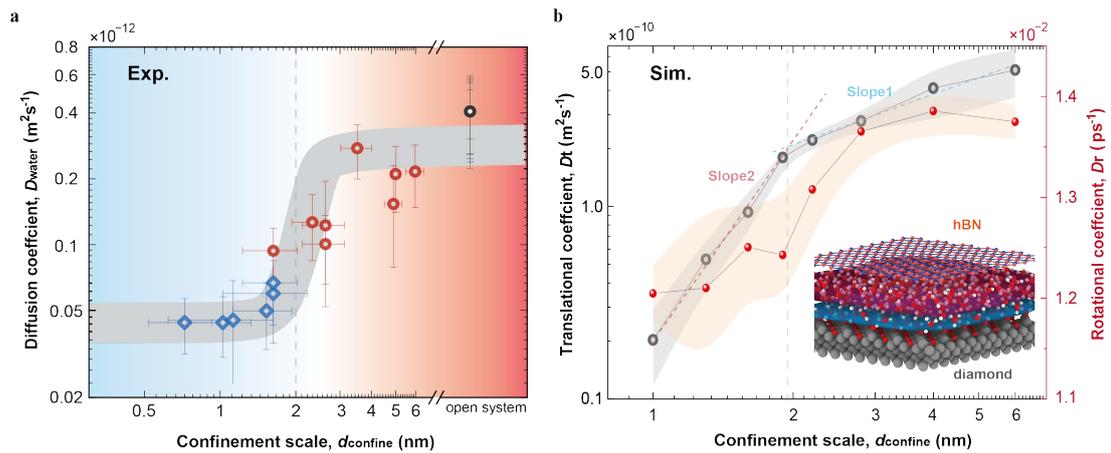

**Fig. 3 | The phase diagram of confined water obtained by experiment and theoretical calculation. (a)** The diffusion coefficient $D_{water}$ under different confinement sizes in log-log scale. The black data points denote $D_{water}$ without hBN confinement, the

red shaded area denotes $D_{water}$ under the weak confinement strength ($d_{confine}$>2 nm), and the blue shaded area denotes $D_{water}$ under the strong confinement strength ($d_{confine}$<2 nm). The error bars of $D_{water}$ arise from the fitting procedure for the correlation spectra of NV as described in Supplementary Text 4. The error bars of $d_{confine}$ arise from the uncertainties during the topographic measurements by AFM. The bold grey curve with the shape of a step function is used for guiding the eye, which demonstrates the liquid-solid phase transition at a critical confinement size of ~2 nm. **(b)** MD simulations of translational (rotational) coefficients $D_t$ ($D_r$) are shown as a grey (red) curve, along with $d_{confine}$ ranging from 1 nm to 6 nm. $D_t$ exhibits a rapid transition from Slope1 (blue dashed line) to Slope2 (red dashed line) near $d_{confine}$~2 nm, and $D_r$ follows a similar trend. The error bars are denoted as shaded areas. Inset: the atomic model of the 2D nanocavity adopted in MD simulations, including the contact water layer (blue part) and bulk-like water (red part) that were confined between the hBN flake and the hydrophilic diamond surface.

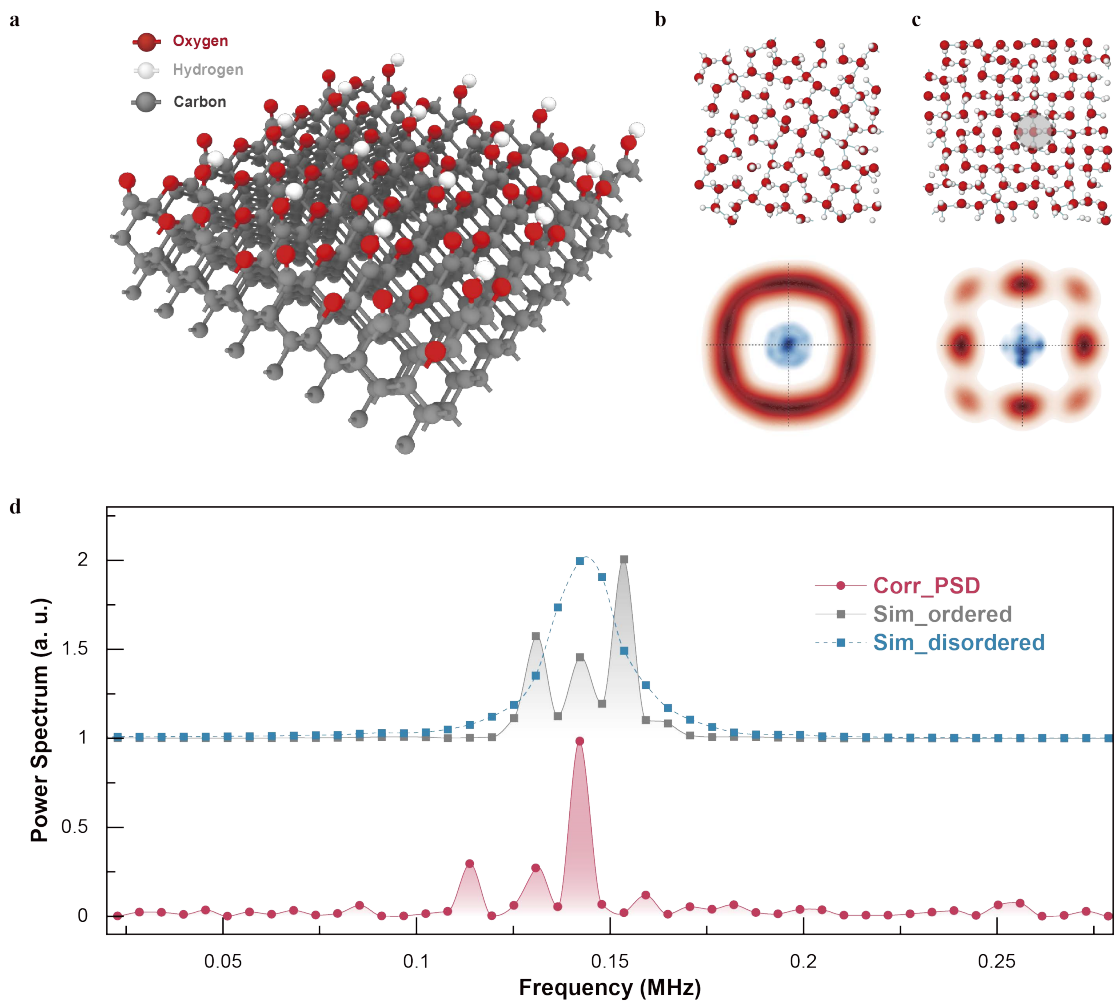

**Fig. 4 | Structure analysis of confined water. (a)** The diamond surface model used in the MD simulations, which includes the ether-like (C-O-C), alcohol-like(C-OH), and ketone-like (C=O) functional groups with the ratio of 2:1:1 to replicate the hydrophilic surface after the acid treatments. **(b)** and **(c)** Upper panel: The hydrogen-bonding networks of the contact water layer under the confinement size of 6 nm and 1 nm, respectively. Lower panel: The corresponding spatial distribution of oxygen (red) and hydrogen (blue) atoms. **(d)** The experimental NMR data in Fig. 2c (red curve) and the calculated one (grey curve) by considering the magnetic dipolar interactions in an averaged representative 5-water cluster (Supplementary Text 7) under the confinement size of 1 nm. A calculated NMR spectrum based on the disordered hydrogen-bonding

network under the confinement size of 6 nm was also shown for comparison (blue dashed curve).

Methods

1. NV-SPM measurement set up.

All the data in this work were recorded in our home-built SPM system[39], which was compatible with the NV-based QS. The SPM part includes a compact Pan-type scanner head, integrated with vector magnets and high-frequency transmission cables. An oil-immersed objective (N. A.=1.3) was used for photon collection, which was driven by a commercial scanner (Physik Instrumente). We chose a qPlus sensor with a tungsten tip (25 µm in diameter) for AFM. The tip was first electrochemically etched in the NaOH solution, and subsequently cleaned and sharpened by focused ion beam (FIB) with a small beam current (typically <80 pA, 30 kV). The spin state of NV was initialized by a 532-nm laser and read out by recording the emitted photons in an APD (Avalanche Photodiode, Excelitas). The 532-nm laser was chopped by an AOM (Acoustic-Optic modulator, Gooch & Housego) crystal in the double-pass mode and subsequently shaped by a single-mode fiber. Microwaves were generated by a Keysight signal generator (N5181B), modulated by an Arbitrary Wave Generator (DN2-662.04, SPECTRUM), where the in-phase and quadrature parts of microwaves were mixed with an external IQ modulator (MMIQ-0218L, Marki). The microwaves were then amplified by a power amplifier, and finally fed through the coplanar waveguide for flipping the electron spin of NV. The synchronization of the laser pulse, microwave pulse, and counter timing was executed through the AWG.

**2. Diamond preparation and generation of shallow NVs.**

The diamond used in this work was an electronic-grade single-crystal chip purchased from Element Six. The intrinsic nitrogen concentration was below 5 ppb. The chips were milled into membranes with a thickness of 20~30 μm by laser cutting in DDK Inc. The diamond membranes were then prepared as described in ref 42. First, an epitaxial boron-doped layer of diamond was grown by microwave-assisted chemical vapor deposition. The thickness of such layer was ~9 nm with a boron concentration above $10^{20}$ /cm$^2$. Next, the diamond was implanted with 5-keV $^{15}$N ions at a tilt angle of 3° to the normal direction of the diamond surface with an average dose of $5\times10^9$ /cm$^2$. A subsequent high-temperature annealing at 950 °C for 2h led to the diffusion of carbon vacancies and the production of NV centers. The sample was then cleaned by boiling in a tri-acid solution. Finally, an ion-etching process by $O_2$ was applied to remove ~15-nm thick diamond in an ICP-RIE (inductively coupled plasma-reactive ion etching) system, resulting in shallow NVs with a depth of ~5 nm.

**3. Fabrication of the 2D nanoconfinement structure.**

hBN flakes were produced by mechanical exfoliation from a single crystal (2D Semiconductors). The flake was firstly exfoliated using a tape (3M Scotch) and then directly transferred onto a PDMS film (WF-X4, Gel-Pak). We skipped the procedure of transferring flakes on a silicon wafer and pick-up by PC stamp in standard transfer technique. This will avoid polymer residuals on the hBN surface that might contribute additional protons signal in NMR. Before the transfer, we characterized several NVs until a good one capable of NMR measurements was found. Then with the help of the markers fabricated on the diamond surface, a multilayer hBN flake (10~20 nm thickness) on PDMS film was selected, navigated with the help of the micro-manipulator (MP285A, Sutter Instruments), and transferred on top of the same NV center (Supplementary Fig. S8). To tune the confinement size, all the fabrication processes were carried out at a controlled temperature and humidity under the protection of nitrogen gas (Supplementary Text 8 and Supplementary Fig. S9).

## 4. Correlation spectrum based on the under-sampling.

For longer delay time in the correlation spectrum, the under-sampling strategy was implemented[50,51]. When the sampling rate $f_s$ is much lower than signal frequency $f_L$ (typically $f_s < f_L/2$), this sampling will introduce an aliasing effect and the measured signal appears to be "folded" into a lower frequency range than the actual one (Supplementary Fig. S4). The observed frequency could be interpreted as $f_L^* = f_L - [f_L/f_s]f_s$, where [] denotes the calculation of the integer quotient. Under-sampling preserves the spectral information like FWHM of each frequency component according to the Nyquist-Shannon criterion[52]. In our case, for the data in Fig. 2c, the Larmor frequency at $f_L$=1.278 MHz was shifted to $f_L^*$=0.1444 MHz under the sampling rate of $f_s$=0.5688 MHz.

**5. Molecular dynamics simulations.**

**Classical MD**. To model the oxygen-terminated diamond (100) surface, we constructed a methoxy-acetone-oxidized diamond (100) surface, as illustrated in Supplementary Fig. S5, which exhibited the lowest formation energy among the predicted atomic structures[53]. Additionally, a diamond (100) surface, with half of the ketone functional groups replaced by hydroxyl groups, was also constructed to simulate the hydrogenated environment, as shown in Fig. 4a. To resemble the pressure induced by the van der Waals interaction between hBN and diamond surface, a finite lateral pressure of ~1 GPa was adopted during the simulation. The calculated diffusion coefficients of the contact water layer on both surfaces revealed a pronounced decrease at a confinement size of ~2 nm, consistent with the experimental observations. The atomic interactions were described using the OPLS all-atom force field and the TIP4P/ice water model for the classical MD simulations[54,55]. All simulations were carried out in the NVT ensemble with a timestep of 1 fs, by utilizing the LAMMPS package[56].

**AIMD**. A diamond (100) surface with an area of 15.1×15.1 Å was modeled to minimize finite size effects, balancing the high computational cost of self-consistent DFT calculations. Water molecules were confined between the diamond and BN surfaces, where the entire BN film and the bottom bilayer carbon atoms of the diamond surface were kept rigid. The vdW-DF2 functional was employed to account for exchange-correlation and weak dispersion interactions[57], while Goedecker–Teter–Hutter (GTH) pseudopotentials and the molecularly optimized double-ζ basis set (DZVP-GTH) were used to describe the core and valence electrons, respectively[58]. Ab initio molecular dynamics (AIMD) simulations were performed for 20 ps with a timestep of 1 fs, by utilizing the CP2k package[59].